\documentclass[10pt, conference, compsocconf]{IEEEtran}
  \usepackage{cite}
 \usepackage{booktabs}

\usepackage{graphicx}
 \usepackage{epstopdf}
  \usepackage{url}
\usepackage[caption = false]{subfig}

\usepackage{multicol}
\usepackage{multirow}
\usepackage{array}

\usepackage{algorithm,algorithmic}

\newcommand{\CASE}[1]{\STATE \textbf{case} #1\textbf{:} \begin{ALC@g}}
\newcommand{\ENDCASE}{\STATE \textbf{end case} \end{ALC@g}}

\newcommand{\PROCEDURE}[2]{\STATE \textbf{procedure} #1(#2) \begin{ALC@g}}
\newcommand{\ENDPRO}{\STATE \textbf{end procedure} \end{ALC@g}}
\newcommand{\FUNCTION}[2]{\STATE \textbf{function} #1(#2) \begin{ALC@g}}
\newcommand{\ENDFUNC}{\STATE \textbf{end function} \end{ALC@g}}

\usepackage{xcolor}
\usepackage{listings}
\usepackage{fancyvrb}
\definecolor{dkgreen}{rgb}{0,0.6,0}
\definecolor{gray}{rgb}{0.5,0.5,0.5}
\definecolor{mauve}{rgb}{0.58,0,0.82}

\begin{document}
\newcommand{\head}[1]{\textnormal{\textbf{#1}}}
\renewcommand\IEEEkeywordsname{Keywords}
%
\title{Towards performance portability through locality-awareness for
  applications using one-sided communication primitives}


\author{\IEEEauthorblockN{Huan Zhou}
\IEEEauthorblockA{High Performance Computing Center Stuttgart (HLRS)\\
University of Stuttgart\\
Germany\\
zhou@hlrs.de}
\and
\IEEEauthorblockN{Jos\'e Gracia}
\IEEEauthorblockA{High Performance Computing Center Stuttgart (HLRS)\\
University of Stuttgart\\
Germany\\
gracia@hlrs.de}
}


%


\maketitle

\begin{abstract}
  MPI is the most widely used data transfer and communication model in
  High Performance Computing. The latest version of the standard,
  \mbox{\mbox{MPI-3}}, allows skilled programmers to exploit all hardware
  capabilities of the latest and future supercomputing systems. The
  revised asynchronous \mbox{remote-memory-access} model in combination with
  the shared-memory window extension, in particular, allow writing
  code that hides communication latencies and optimizes communication
  paths according to the locality of data origin and destination. The
  latter is particularly important for today's multi- and many-core
  systems. However, writing such efficient code is highly complex and
  error-prone. In this paper we evaluate a recent \mbox{remote-memory-access}
  model, namely \mbox{DART-MPI}. This model claims to hide the
  aforementioned complexities from the programmer, but deliver
  locality-aware \mbox{remote-memory-access} semantics which outperforms
  \mbox{MPI-3} one-sided communication primitives on multi-core systems.
  Conceptually, the \mbox{DART-MPI} interface is simple; at the same time it
  takes care of the complexities of the underlying \mbox{MPI-3} and system
  topology. This makes \mbox{DART-MPI} an interesting candidate for porting
  legacy applications. We evaluate these claims using a realistic
  scientific application, specifically a finite-difference stencil code which
  solves the heat diffusion equation, on a large-scale Cray XC40
  installation. 

\end{abstract}

\begin{IEEEkeywords}
DART-MPI; one-sided; application porting; data-locality
\end{IEEEkeywords}

%
\IEEEpeerreviewmaketitle

\section{Introduction}

Numerical simulation using distributed and parallel applications, 
e.g., studying aerodynamics, fluid dynamics, molecular dynamics,
weather forecasting and so on,
provides researchers with insight into complex natural  phenomena.
Various parallel programming models are developed for implementing
efficient and scalable distributed simulations.
The Message Passing Interface (MPI,~\cite{mpi3-standard}) remains as the dominant communication model as a result of
its high performance, portability and standardization on cutting-edge
computing systems.

Multi- or many-core processors are commonly deployed in 
today's computation clusters due to it fuels
an explosive growth of processing capability.
According to the new TOP500 Supercomputer report~\cite{top500},
the most powerful supercomputer as of June 2016 -- Sunway --
exemplifies this fact with core counts exceeding 10,000,000.
However, taking into account additional resource sharing,
communication and synchronization  within a node of 
multi- and many-core clusters
poses new challenge for design of parallel applications.
To get the optimal intra-node performance,
researchers need to design their parallel programs to be aware of
data-locality and exploit all available communication paths. 
For instance, data exchange within a node should bypass the MPI 
communication primitives and use direct memory load/store operations.

The MPI standard has been evolving
to keep up with the scalable and productive computation
hardware. MPI-2 incorporates the one-sided interfaces to
avoid the matching affairs inherent in the two-sided operation.
Moreover, in \mbox{MPI-3} \mbox{remote-memory-access} (RMA) is extended
to support shared memory windows.  The
results provided in earlier works~\cite{conf/pvm/Hoefler, conf/openshmem/Hammond}
have proved that the
MPI-integrated shared memory window is a promising alternative
since it allows co-existence of RMA operations and  direct load/store accesses.
However, note that MPI does
not intuitively expose data locality to user applications.
Thus, the programmer needs to have profound knowledge of the details
in MPI to be able to write efficient, data-locality aware applications 
based on MPI. %
In addition, the MPI interfaces for RMA and shared-memory extensions are
very complex, making the development error-prone and time-consuming.

\mbox{DART-MPI}~\cite{conf/pgas/ZhouMIGGF14} is an \mbox{MPI-3} based implementation
of the runtime system for the hierarchical PGAS-like C++ programming model
DASH~\cite{dash}. However, in this paper \mbox{DART-MPI} is treated as a
programming model in its own right. 
It provides a number of communication primitives including
one-sided and collective operations through a plain C-based interface.
One fundamental goal of \mbox{DART-MPI} is to support 
data-locality aware communication schemes while
keeping the user away from MPI complexities. 
In particular, \mbox{DART-MPI} internally takes responsibility over all those details
related to the correctness and performance mentioned above.
Therefore, \mbox{DART-MPI} enables portable and efficient
parallel programming for todays multi- and many-core
clusters. Besides,
the DART RMA communication performance penalties for wrapping
the underlying MPI-3 and system topology are shown in 
the dissertation~\cite{diss/zhou} and they become negligible when
the larger messages are transferred.

We briefly discuss some background used in this paper in section
\ref{sec:background}, explain how \mbox{DART-MPI} hides detail and
complexities of \mbox{MPI-3} RMA from the user and demonstrate
the terseness and clarity of \mbox{DART-MPI} code
in comparison to locality aware \mbox{MPI-3} code in
section~\ref{sec:DART-code}. Finally, in section~\ref{sec:eval} we
demonstrate the performance of \mbox{DART-MPI} using a heat diffusion problem,
which is a representation for a wide class of numerical simulation
schemes. %
In particular, we present a comprehensive picture of the comparative
results between the \mbox{DART-MPI} and \mbox{MPI-3} RMA implementations of this
application on large-scale, supercomputers such as Cray XC40 systems.

\section{Background}
\label{sec:background}
\subsection{\mbox{DART-MPI}}

DART~\cite{conf/pgas/ZhouMIGGF14} defines common concepts, terminology
and abstracts from the underlying communication substrate
and hardware.
DART establishes a partitioned global address space 
and provides functions to handle memory efficiently,
such as memory allocation and data movement.
In addition, DART also provides functions for
initialization, synchronization and management of teams,
which are similar to the MPI communicators.
In  a  DART  program  an individual participate is called 
unit. DART provides the functions of {\em dart\_init} and {\em dart\_exit}
for initialization and finalization.
Importantly, providing and working with a global memory (collective or non-collective) is the focus
of DART.
The global address space is a virtual abstraction, with each
unit contributing a part of its local memory. Remote data items are
addressed by global pointers provided by DART.
The complete DART specification is available on-line at
\url{https://dl.dropboxusercontent.com/u/408013/dart\_spec\_v2.1.html}.

When possible
(i.e., intra-node),
the \textit{shmem-win} (shared memory window object) described in the 
paper~\cite{conf/europar/ZhouIG15}
are always used by \mbox{DART-MPI} blocking RMA operations to 
perform load or store instructions.
Additionally, the \mbox{DART-MPI} intra-node non-blocking RMA operations are 
designed as
the MPI RMA operations on \textit{shmem-win},
whilst all the \mbox{DART-MPI} inter-node RMA operations
turn to the MPI RMA operations on \textit{d-win} (MPI dynamically-created window object)
or \textit{win} (MPI-created window object).
Hence we can observe that
\mbox{DART-MPI} spontaneously has data locality in mind
when performing the RMA operations.
DART blocking and non-blocking RMA operations are represented as
{\em dart\_get/put\_blocking} and {\em dart\_get/put}, respectively.


\subsection{Heat conduction}
\begin{figure}
\begin{center}
\includegraphics[width=0.9\linewidth]{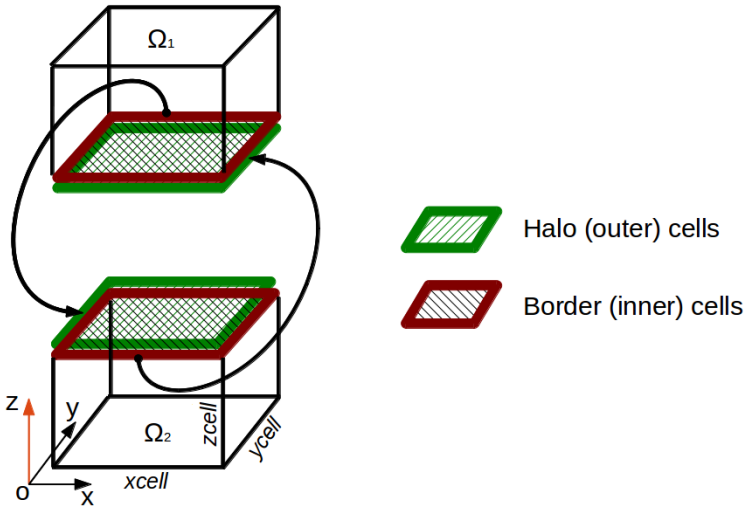}
\caption[Halo cells exchange in 3D grid.]
{Halo cells exchange in 3D grid.
Exchange of the matrix data 
(a total of \textit{xcell}$\times$\textit{ycell} grid cells) happen
on Oxy plane between neighboring
processes in the direction of z-axis,
where sub-domain $\Omega_{2}$ is the 
\textit{Down} neighbor of sub-domain $\Omega_{1}$.}
\label{3dheatexchange}
\end{center}
\end{figure}


The heat conduction\cite{heatconduction} is a mode of heat transfer owing to
molecular activity and occurs in any material (e.g., solids, fluids and gases).
We used the application source code parallelized with MPI at~\cite{3Dheat}.
This application simulates the phenomenon of 3D heat conduction in solids with
temperature-dependent thermal diffusivity.
The Partial Differential Equation (PDE) for 
\textit{unsteady} 3D heat conduction is obtained in 
the Cartesian coordinate system.
Here, \textit{unsteady} means the temperatures 
may change during the process of heat conduction.
This application solves the PDE over a 3D grid by 
using the Finite Difference Method (FDM).
Boundary conditions are constant temperatures at the 
edges of the 3D grid.

Moreover,
parallelization is done based on the checkerboard domain decomposition.
After the decomposition
each process has six neighbors located in the direction
of \textit{East-West} (x-axis),
\textit{North-South} (y-axis) and \textit{Up-Down}
(z-axis) according to the Cartesian process
topology. The exceptional case is the processes owning
the edge cells will not have all of the six neighbors.
Each process sends each face of its sub-domain (a 3D sub-grid), i.e.,
matrix, to its corresponding neighbor for the computation.

Each process exchanges the border data with its 
corresponding neighbors and the halo cells are reserved 
to receive the exchanged data.
The original 3D heat conduction application fills the halo cells
in each iteration using the point-to-point communication
operation (i.e., {\em MPI\_Sendrecv}).
Each cell of the grid is a 8-byte double-precision floating-point number.
The abort-criterion convergence is achieved by invoking
the collective operation, such as all-to-all reduce.
Figure~\ref{3dheatexchange}, 
takes the direction of z-axis for example,
to explain the way of boundary
data exchange between two neighboring processes.

\section{Method of wrapping MPI complexity inside DART interfaces}
\label{sec:DART-code}
\begin{figure}[h]
 \begin{lstlisting}[mathescape, escapechar=ä]
  MPI_Init ($\ldots$);
  MPI_Comm_size (MPI_COMM_WORLD, &size);
  ä\textcolor{red}{MPI\_Comm\_split\_type}ä
  ä\textcolor{red}{(MPI\_COMM\_WORLD,$\ldots$, \&sharedmem\_comm);}ä
  MPI_Win_create_dynamic (MPI_COMM_WORLD, &d-win);
  MPI_Win_lock_all (d-win);
  ä\textcolor{red}{if (sharedmem\_comm!=MPI\_COMM\_NULL)\{}ä
     ä\textcolor{red}{MPI\_Win\_allocate\_shared}ä
     ä\textcolor{red}{(nbytes,$\ldots$, \&membase, \&sharedmem\_win);}ä
     ä\textcolor{red}{MPI\_Win\_lock\_all (sharedmem\_win);\}}ä
  MPI_Win_attach (d-win, membase, nbytes);
  MPI_Get_address (membase, &disp);
  disp_s = (MPI_Aint*)malloc (size * (sizeof(MPI_Aint)));
  MPI_Allgather (&disp, 1, MPI_AINT, disp_s,$\ldots$, MPI_COMM_WOLRD);
  dest = randomdst_generator ();
  /* The array is_shmem indicates the position of
  the given target process with respect to the origin */
  ä\textcolor{red}{if ((j=is\_shmem[dest])>=0)\{}ä//on-node communication
     /* The j is the relative target rank in the 
     corresponding sharedmem_comm */
     ä\textcolor{red}{MPI\_Win\_shared\_query}ä
     ä\textcolor{red}{(sharedmem\_win, j,$\ldots$, \&baseptr);}ä
     ä\textcolor{red}{memcpy (baseptr, srcptr, nbytes);\}}ä
  else{//across-node communication
     MPI_Put (srcptr, nbytes, dest, disp_s[dest],$\ldots$, d-win);
     MPI_Win_flush (dest, d-win);}
  MPI_Win_detach (d-win, membase);
  ä\textcolor{red}{if (sharedmem\_comm!=MPI\_COMM\_NULL)\{}ä
     ä\textcolor{red}{MPI\_Win\_unlock\_all (sharedmem\_win);\}}ä
  MPI_Win_unlock_all (d-win);
  ä\textcolor{red}{MPI\_Win\_free (\&sharedmem\_win);}ä
  MPI_Win_free (&d-win);
  free (disp_s);
  MPI_Finalize ();
\end{lstlisting}
\caption[An example for MPI code.]{An example for MPI code.
This example shows the MPI code for transferring data from 
an origin process to another random target process with
data locality in mind.}
\label{mpicodeexample}
\end{figure}

\begin{figure}[h]
 \begin{lstlisting}[mathescape, escapechar=']
  dart_init ();
  dart_team_memalloc_aligned 
  (DART_TEAM_ALL, nbytes, &gptr);
  dest = randomdst_generator ();
  dart_gptr_setunit (&gptr, dest);
  dart_put_blocking (gptr, srcptr, nbytes);
  dart_team_memfree (DART_TEAM_ALL, gptr);
  dart_exit ();
\end{lstlisting}
\caption[An example for DART code.]{An example for DART code.
This example shows the DART code for transferring data from 
an origin unit to another random target unit with
data locality in mind.}
\label{dartcodeexample}
\end{figure}

Figure~\ref{mpicodeexample} shows the MPI code that is 
used to consciously send a message from the origin to another
random target process with the data locality in mind. It is clearly
observed that in MPI code, the window creation/destroy operations
as well as the access epoch start/end operations should be dealt
with explicitly and carefully. In detail,
the MPI programmer first needs to 
allocate a shared-memory window along with a region of 
memory with specified size and then creates another
RMA window spanning the same region of memory.
Besides the creation of 
these two nested windows, a hash table (named as is\_shmem)
is entailed to indicate the data locality.
Finally, different communication paths are taken according to
the data locality information (on-node or across-node).
Regarding the on-node communications, direct load or write accesses are allowed.
Otherwise, the MPI RMA communication operations are employed.
Figure~\ref{dartcodeexample} shows the corresponding data-locality
aware code of DART.
Revisiting Fig.~\ref{mpicodeexample}, the red code lines are required
for handling the intra-node data transfers.
In comparison to the 
MPI code, the DART code
is obviously more concise and easier-to-read.
This is due to that
\mbox{DART-MPI} has internally taken over
the responsibility for the 
locality-awareness and manipulation of access epoch 
by hiding the complexity from the user.

\begin{figure}[h]
 \begin{center}
  \includegraphics[width=\linewidth]{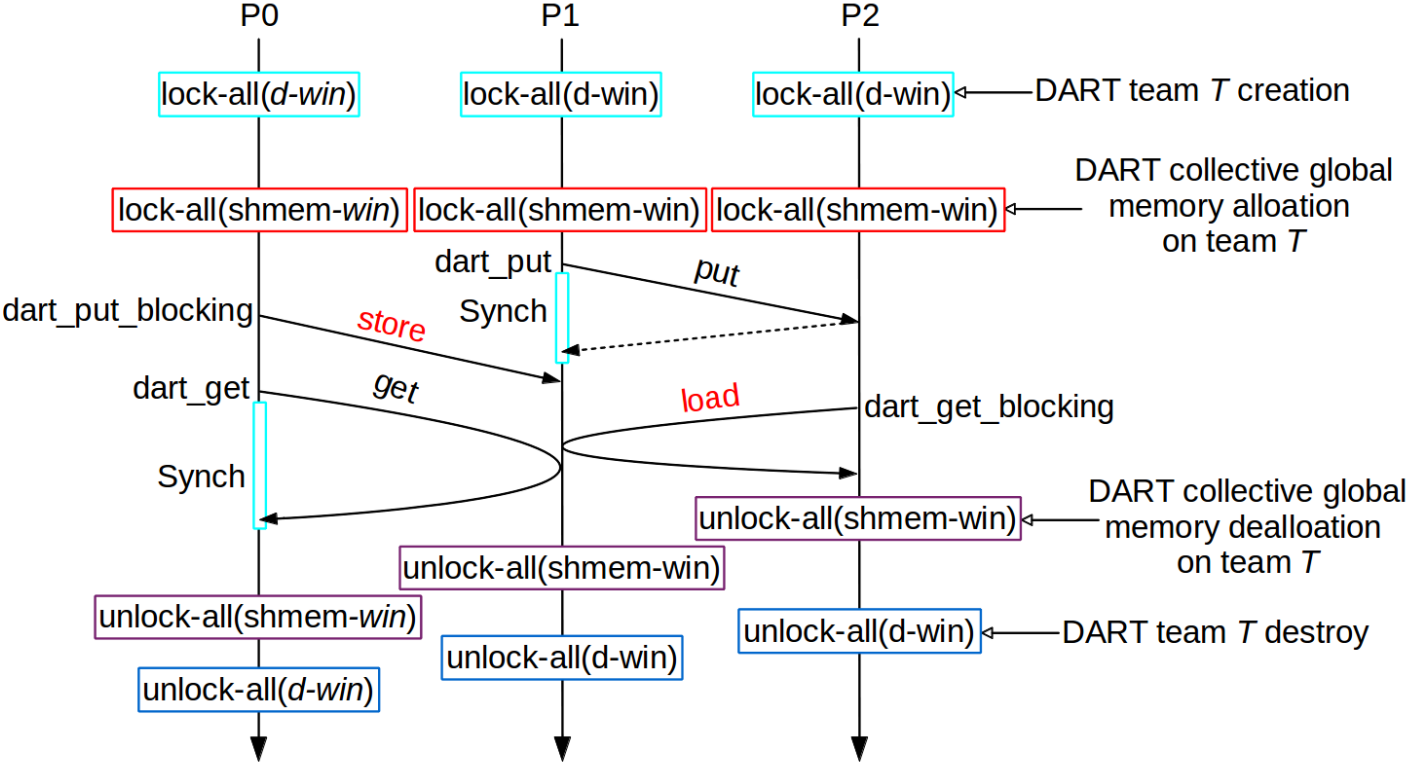}
 \end{center}
 \caption[Example of DART RMA operations within two passive epochs.]
 {Example of \mbox{DART-MPI} RMA operations within two passive epochs.
 Here, the two passive epochs are started by a call to {\em MPI\_Win\_lock\_all}
 and ended by a call to {\em MPI\_Win\_unlock\_all}.
 The Synch means the operation of {\em dart\_wait}.}
 \label{lockall}
\end{figure}

The MPI RMA communication operations coupled with explicit 
passive synchronization operations
(with shared lock) form a theoretic foundation for building 
the \mbox{DART-MPI} RMA model~\cite{conf/pgas/ZhouMIGGF14}.
In fact, each MPI RMA communication call
must proceed within an access epoch for this
window at a process to start and 
complete the issued RMA communications.
Superficially, the RMA semantics in MPI and DART
both require the independent management of
synchronization and communication. However,
the concept of access epoch is not relevant in
the DART RMA semantics according to the DART specification.
Therefore, all primitives pertaining to MPI access
epoch need to be hidden from DART programmer. Importantly,
this focus creates the illusion that all DART global memory
regions are protected and exposed once allocated and 
also the expected processes are allowed to access
those global memory regions directly.


\mbox{MPI-3} supports a global lock mechanism providing the 
lock\_all/unlock\_all routines (i.e., {\em MPI\_Win\_lock\_all}
/{\em MPI\_Win\_unlock\_all}) to allow
locking/unlocking multiple targets simultaneously.
I.e., 
the pair of {\em MPI\_Win\_lock\_all} and {\em MPI\_Win\_unlock\_all}
enables programmer to lock/unlock all processes in a certain window
with a shared lock. 
This mechanism avoids the proneness to
error and repeated open or closure operations on the access epoch.
Thus, it is applied to build up
\mbox{DART-MPI} RMA model by leveraging 
the global lock/unlock routines.
Given the DART collective and non-collective 
global memory managements differ in concept and design,
we will explain how
to safely expose these two types of global memory
independently based on the global lock mechanism.

A global \textit{win}
and a global \textit{shmem-win}
are created once a DART program is initiated~\cite{conf/europar/ZhouIG15}.
Therefore, 
each process/unit needs to add two {\em MPI\_Win\_lock\_all} calls
for the global \textit{win} and \textit{shmem-win} to
start two protected shared access epochs to all other units.
Prior to freeing up the global \textit{win}
and \textit{shmem-win} (done inside {\em dart\_exit}),
each unit issues two matched {\em MPI\_Win\_unlock\_all} calls
to end the above two access epochs.

Unlike non-collective global memory allocations,
collective global memory allocations always involve
all units in the given team.
Thus,
there are two scenarios where the creation of \mbox{\textit{d-win}}~\cite{conf/europar/ZhouIG15}
takes place:
\begin{enumerate}
 \item When a DART program is launched, 
a \textit{d-win} for {\em DART\_TEAM\_ALL} is created.
 \item When a new team (e.g., team $T$) is created,
a \textit{d-win} for team $T$ is created.
\end{enumerate}
In each of the above scenarios, 
a shared lock all epoch for the \textit{d-win} is started.
Additionally,
we associate a \textit{shmem-win} with
a region of allocated collective global memory.
This necessitates the call to 
{\em MPI\_Win\_lock\_all} for the \textit{shmem-win}.

Likewise, 
we can complete the lock all epoch by
a call to {\em MPI\_Win\_unlock\_all}
when the DART program is finalized or 
team $T$ is destroyed.
A call to {\em MPI\_Win\_unlock\_all} 
is also entailed to complete a shared RMA access epoch
along with the deallocation of the collective 
global memory region.
Remote completion
of communications
without ending the access epoch can be achieved with
the \textit{flush} routines (i.e., {\em MPI\_Win\_flush} and its all-, any- and 
local-variants).
An {\em MPI\_Win\_flush} specifies
a certain target and ensures that all 
previous operations to the target are 
locally and remotely finished.

The DART synchronization routines ({\em dart\_wait} and {\em dart\_waitall})
are expected to ensure the 
remote and local completion of
DART non-blocking RMA communications.
Therefore, the explicit MPI bulk synchronization
using {\em flush} is integrated into
{\em dart\_wait}.
Likewise,
the {\em dart\_waitall} performs a series of 
related calls to {\em MPI\_Win\_flush}.
Figure~\ref{lockall} thoroughly 
shows the exemplary DART RMA
events happening on the collective
global memory region across team $T$ with MPI global lock mechanism.
Here  we assume that the team $T$ consists of three processes locating within one node.

\section{Experimental evaluation}
\label{sec:eval}
\begin{figure}[ht!]
 \begin{lstlisting}[mathescape, escapechar=']
  /* Communicate matrix data on Oxz plane */
  if (South neighbor exists)
    for (i = 0; i < $xcell$; i++)
      get ($zcell$, South); // Get zcell consecutive grid cells from South neighbor
  barrier;
  if (North neighbor exists)
    for (i = 0; i < $xcell$; i++)
      get ($zcell$, North);
  barrier;
  /* Communicate matrix data on Oyz plane */
  if (West neighbor exists)
    for (i = 0; i < $ycell$; i++)
      get ($zcell$, West); // Get zcell consecutive grid cells from West neighbor
  barrier;
  if (East neighbor exists)
    for (i = 0; i < $ycell$; i++)
      get ($zcell$, East);
  barrier;
  /* Communicate matrix data on Oxy plane */
  if (Up neighbor exists)
    for (i = 0; i < $xcell$; i++)
      for (j = 0; j < $ycell$; j++)
	       get ($1$, Up); // Get one grid cell from Up neighbor
  barrier;
  if (Down neighbor exists)
    for (i = 0; i < $xcell$; i++)
      for (j = 0; j < $ycell$; j++)
	       get ($1$, Down);
  barrier;
\end{lstlisting}
\caption{Pseudo-code for the halo cells exchange with get and barrier operations.}
\label{3dheatcode}
\end{figure}
In this section we present a comprehensive picture
of the comparative results
between \mbox{DART-MPI} RMA and MPI RMA through
a numerical simulation of 3D heat conduction problem.
In this experiment,
we stop the calculation after 5000 iterations
and collect and analyze the time results
of the computation 
(update the inner grid cells)
and halo cells exchange
for different grid sizes
and participating processes.
We plot the average data results of 25 runs with 
small execution time variation reported.
\subsection{Experimental testbed}
The experimental environment was the Cray XC40 system~\cite{hlrswiki}.
Cray XC40 system is equipped with 7,712 compute nodes
made up of dual twelve-core Intel Haswell
E5-2680v3 processor (one processor per socket), which has 
exclusive 256 KB L2 unified cache for each core.
Therefore, each compute node has
24 cores running at 2.5GHZ with 128 GB of DDR4 (Double 
Data Rate) RAM (Random Access Memory).
The different compute nodes are interconnected
via a Cray Aries network using Dragonfly topology.
The Cray XC40 system features a hierarchical
network topology.
Detailedly, the Cray XC Rank-1 network
is used for communications happening across
different compute blades within a chassis
over backplane. Each chassis consists of 16 compute blades.
Communications happening across different chassis
within a 2-cabinet group over copper cables
are supported with
the Cray XC Rank-2 network.
Further, the Cray XC Rank-3 network is employed
for communications happening between the groups
over the optical cables.
All the experiments in this paper are
based on the Cray Programming Environment
5.2.56.
Moreover, the tasks are assigned to cores in SMP style~\cite{rankplacement}, which
means one node needs to be filled before going to next.

\subsection{\mbox{DART-MPI} and MPI RMA port of 3D heat conduction application}

For our test, we first implement the above mentioned 3D heat conduction
algorithm based on MPI one-sided interfaces using passive target mode.
We then port it to DART one-sided directives, where just several ($\sim$ 14)
modified code lines are required.
The porting procedure is thus straightforward and economic.
The halo cells exchange is achieved by
using blocking get operation. We can 
deduce that six blocking get operations
will be invoked for each process.
Particularly, {\em MPI\_Rget} is invoked first and then
followed by {\em MPI\_Wait} in MPI implementation and 
{\em dart\_get\_blocking} is used in DART implementation.
Note that, within each calculation iteration,
after one round of halo cells exchange with certain neighbor
an explicit process synchronization, 
such as barrier ({\em MPI\_Barrier} in MPI
and {\em dart\_barrier} in \mbox{DART-MPI}),
is naively inserted in our code
due to an implicit synchronization is ideally
supported by point-to-point communication model.
Therefore, the halo cells exchange time
measured below includes the overhead brought
by process synchronization. 
Regarding the process synchronization, we only measure
the overhead introduced by the barrier operations instead of
the time to wait between processes.
Figure~\ref{3dheatcode} concisely shows the critical part
of the halo cells exchange code within each calculation
iteration.


\subsection{Performance results}
\begin{figure*}[!ht]
\begin{center}
\includegraphics[width=0.9\linewidth,height=0.31\textheight]{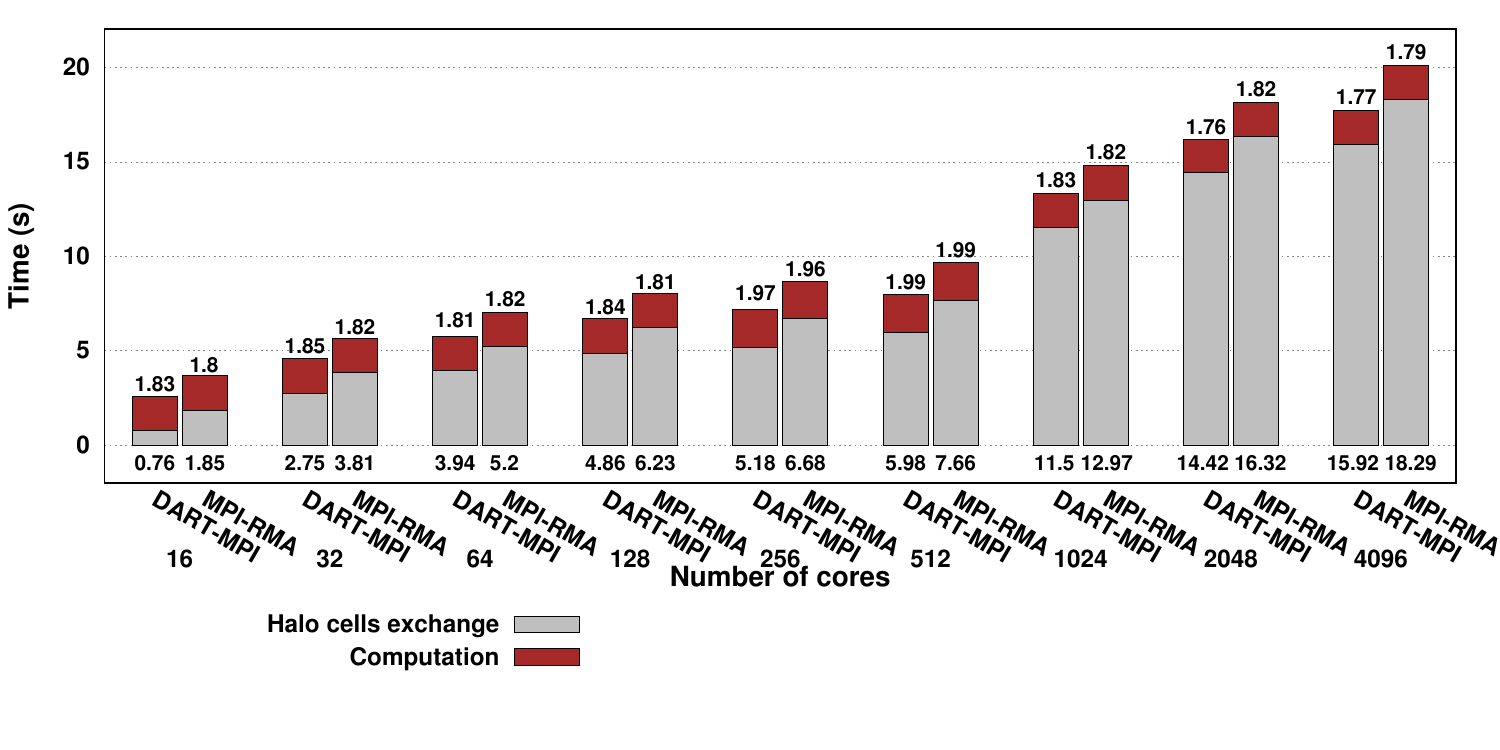}
\caption[Weak scaling for one-sided DART and MPI.]
{Weak scaling for one-sided DART and MPI. The grey bar signifies the halo cells exchange
time, the value of which is  
written on the bottom.
The brown bar signifies the 
computation overhead,
the value of which
is written on the top.}
\label{weakscale32KB}
\end{center}
\end{figure*}

\begin{figure*}[!ht]
\begin{center}
\includegraphics[width=0.9\linewidth,height=0.31\textheight]{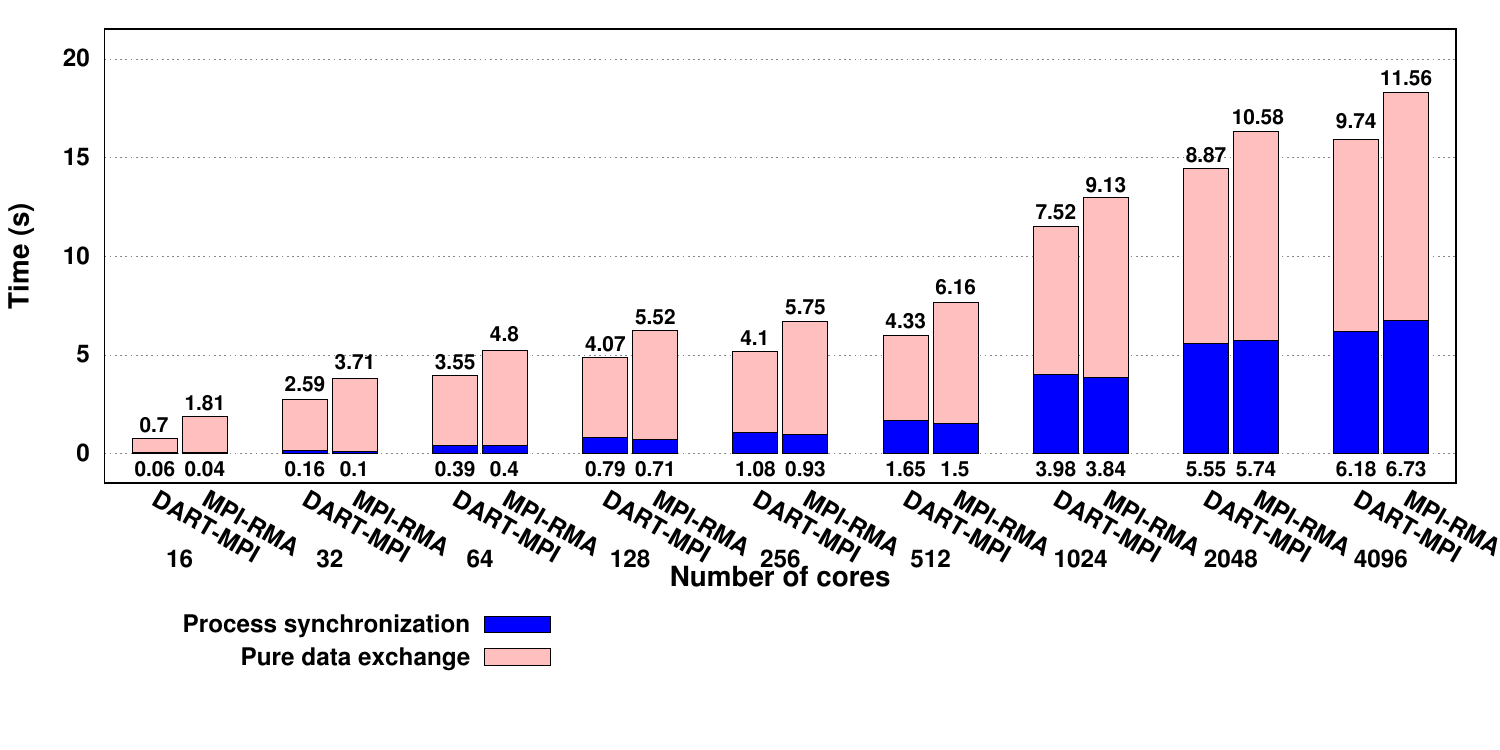}
\caption[
Breakdown of halo cells
exchange time for
the weak scaling problem.]
{Breakdown of halo cells
exchange time for
the weak scaling problem.
The blue bar signifies the overall
process synchronization overhead, the value of which
is written on the bottom.
The pink bar signifies the pure data
exchange overhead, the value of which
is written on the top.
The sum 
equals to the halo cells exchange overhead
that we measured in this benchmark. 
}
\label{weakscale32KBcomm}
\end{center}
\end{figure*}

Figure~\ref{weakscale32KB}
gives the quantitative results showing the 
performance of \mbox{DART-MPI} RMA and MPI RMA on Cray XC40
system over the number of cores (from 16 to 4096).
A weak scaling evaluation, where 
the number of grid cells per core is fixed and
is applied with grid sizes
varying from 
(64$\times$64$\times$128) to (256$\times$2048$\times$256).
The domain size is chosen such, that overall execution is dominated by
the time to exchange halo cells in order to increase the statistical
significance of the latter. 
As expected from a weak scaling experiment, the total execution time
increases only slowly due to larger relative communication cost. In fact, the actual
computation time, i.e. the time to update the inner grid cells, stays
nearly constant with the increasing core count and is the same for the MPI RMA
and \mbox{DART-MPI} version within the observed statistical variance. 
However, the \mbox{DART-MPI} implementation can provide 
a relative speedup of $\sim$1.38x over MPI
on average in terms of
the halo cells exchange time performance.
Such speedup is mostly attributed to
the enabling of direct load/store access
within one node in the \mbox{DART-MPI}.
Particularly,
when only the on-node performance is considered,
that is 16 cores, 
then the improvement can be obviously seen,
\textit{cf.} 0.76s for \mbox{DART-MPI}, and 1.85s for MPI.
In addition,
the consistent improvement seen is expected because most of 
the data exchanges are still local to a node
in this evaluation.
Such fact highlights
and visualizes the advantage
of the memory sharing mechanism in intra-node case
employed by \mbox{DART-MPI} implementation.
Furthermore, we can observe that
the performance speedup gets
decreased as the number of cores is increased,
which is not surprising given the number
of across-node data exchanges is accordingly
increased and DART RMA uses the same communication
infrastructure as MPI RMA internally
in the inter-node case.

On the other hand,
shown in Fig.~\ref{weakscale32KB},
the halo cells exchange
time gets sustained growth over the number 
of cores.
Specifically, 
a sudden rise occurs when the communications
go beyond one node and start crossing nodes (32 cores).
The fact, that is,
several inter-node data transfers
start to be involved accordingly, 
creates big difference in performance.
Besides that, 
the process synchronization may
also be a part of the growth
of the halo cells exchange time,
as hinted above.
Therefore, Figure~\ref{weakscale32KBcomm}
breaks down the halo cells exchange time
in terms of pure data exchange and process
synchronization, which can help us understand
how the two components affect the 
amount of time the halo cells exchange takes
to run.
Observing this figure, it is immediately clear
that both the process synchronization and pure data exchange
overhead increase gradually over the number of cores.
The breakdown information
tells us that the pure data exchange component plays 
an important role in the halo cells exchange
time rise especially 
when the number of cores reaches up to 1024.
In this experiment, inter-group communications through
Cray XC Rank-3 optical network take place when 1024
cores are launched, which would greatly degrade
data exchange performance, in comparison to the inter-blade (intra-chassis)
communications through
Cray XC Rank-1 backplane network and inter-chassis (intra-group) communications
through Cray XC Rank-2 
copper network.
This is the reason for the sudden rise in the 
pure data exchange time when ranging the number
of cores from 512 to 1024.

Here, we would add that 
the Fig.~\ref{weakscale32KBcomm}
sufficiently illustrates that
the data-locality aware RMA operations
in the \mbox{DART-MPI} result in the improvement
in the halo cells exchange time.

\section{Related work}
A heuristic study~\cite{RMAcomp} conducted on Cray XE6
shows that one-sided MPI-2 has relatively
inferior scaling behavior compared to UPC and 
Cray SHMEM RMA.
This is due to that random allocated memory can be specified
to be remotely accessible in MPI-2.
Unlike MPI-2, the memory for RMA operation is somehow
customized for Cray SHMEM (symmetric memory) and UPC (shared memory).
Further, \mbox{MPI-3} RMA enables direct load/store accesses by supporting
a shared-memory window~\cite{conf/pvm/Hoefler, conf/openshmem/Hammond},
which is employed in \mbox{DART-MPI} to 
enhance the intra-node RMA communication performance.

Additionally, MPI researchers conduct 
long-term optimization works on the
MPI RMA~\cite{conf/ipps/LiuJWPABGT04, ping, 
conf/sc/GerstenbergerBH13, conf/pvm/PotluriWDSP11}
and 
collective operations~\cite{mpich-opt,
conf/ipps/LiuMP04, conf/ccgrid/MamidalaKDP08, lmbcast}.
Those optimizations are also liable to
be applied onto \mbox{DART-MPI} and then benefit
the performance of applications.

\section{Conclusion}
We depicted the approach of leveraging the existing MPI
RMA synchronization interfaces (global lock mechanism)
for the related DART interfaces. The DART code of data
locality-awareness is more concise and easier-to-read than
MPI code. This is due to that \mbox{DART-MPI} has internally
taken over the responsibility for the locality-awareness and 
manipulation of epoch access. We then described
the rewriting structures for porting a 3D heat conduction
application onto one-sided MPI and then onto 
\mbox{DART-MPI}. The \mbox{DART-MPI} port
is proved to be efficient for different number of
cores and grid sizes on Cray XC40 system and speeds up the
communication performance (halo cells exchange) by 27\%
on average over MPI port in this application level
evaluation. Such performance improvement is induced
by the fact that one-sided DART is implemented using the
feature of \mbox{MPI-3} integrated shared memory window in the
intra-node case while retaining the MPI inter-node RMA
performance. Therefore, \mbox{DART-MPI} has basically reached
the goal of making itself easier for programmers to write
efficient applications with the data locality in mind. Further,
the \mbox{DART-MPI} applications can be executed on 
different computers with hierarchical memory architecture 
(like Cray XC40 system) without any code change. Meantime,
the performance advantages brought
by the factor of locality awareness will also be retained
since the underlying MPI communication conduit is highly supported 
and tuned by a wide range of hardware systems.


\section*{Acknowledgment}
This work has been supported by the European Community
through the project Polca (FP7 programme under
grant agreement number 610686) and Mont Blanc 3 (H2020
programme under grant agreement number 671697). We
gratefully acknowledge funding by the German Research
Foundation (DFG) through the German Priority Programme
1648 Software for Exascale Computing (SPPEXA) and the
project DASH.



%
\bibliographystyle{IEEEtran}
\bibliography{IEEEabrv,./paper}

\end{document}